\documentclass[12pt]{article}

\usepackage{graphicx}
\usepackage{caption} 
\usepackage{subcaption}  

\newcommand{\bbeta}{ \mbox{\boldmath $\beta$} }

\newcommand{\beps}{ \mbox{\boldmath $\epsilon$} }

\newcommand{\bA}{ \mbox{\bf A} }
\newcommand{\bB}{ \mbox{\bf B} }

\newcommand{\bD}{ \mbox{\bf D} }

\newcommand{\bF}{ \mbox{\bf F} }
\newcommand{\bG}{ \mbox{\bf G} }
\newcommand{\bh}{ \mbox{\bf h} }
\newcommand{\bH}{ \mbox{\bf H} }
\newcommand{\bI}{ \mbox{\bf I} }

\newcommand{\bL}{ \mbox{\bf L} }

\newcommand{\bR}{ \mbox{\bf R} }

\newcommand{\bt}{ \mbox{\bf t} }

\newcommand{\bu}{ \mbox{\bf u} }

\newcommand{\bW}{ \mbox{\bf W} }

\newcommand{\bX}{ \mbox{\bf X} }

\newcommand{\by}{ \mbox{\bf y} }

\newcommand{\bZ}{ \mbox{\bf Z} }
\newcommand{\bze}{ \mbox{\bf 0} }
\newcommand{\bone}{ \mbox{\bf 1} }

\begin{document}

\begin{center}
{\Large \textbf{Statistical methods research done as science rather than mathematics}}

\vspace{10pt}
\textbf{James S. Hodges}\\
\vspace{10pt}
Division of Biostatistics, University of Minnesota, Minneapolis,
Minnesota USA 55414\\
\emph{email:} hodge003@umn.edu\\
\vspace{20pt}

\today
\vspace{20pt}

\textsc{ABSTRACT}
\end{center}
\vspace*{-10pt}
This paper is about the way we study statistical methods.  As an example, it uses the random regressions model, in which data come in clusters and the intercept and slope of cluster-specific regression lines are treated as a bivariate random effect.  Maximizing this model's restricted likelihood is prone to produce an estimate of $-1$ or +1 for the correlation between the two random effects or of 0 for one of the random-effect variances.  We argue that this is a problem;  that the problem lacks an explanation, not to mention a solution, because our discipline has developed little understanding of how contemporary models and methods map data into inferential summaries;  that such understanding is absent, even for a model as simple as random regressions, because of a near-exclusive reliance on mathematics as a tool to gain understanding;  and that judging from our literature, math alone is no longer sufficient for this task.  We then argue that as a discipline, we can and should add a tool to our toolkit, breaking open our black-box methods by mimicking the five steps that our colleagues in molecular biology commonly use to break open Nature's black boxes:  design a simple model system, formulate hypotheses using that system, test them in experiments on that system, iterate as needed to reformulate and test hypotheses, and finally test the results in an ``{\it in vivo}" system.  We demonstrate this style of inquiry by using it to understand conditions under which the random-regressions restricted likelihood is likely to be maximized at a boundary value.  Resistance to this empirical approach to gaining understanding seems to arise from a view that it lacks the certainty or intellectual heft of mathematics, perhaps because simulation experiments in our literature rarely do more than measure  a new method's operating characteristics in a small range of situations.  We argue that such work can make useful contributions including, as in molecular biology, the findings themselves and sometimes the designs used in the five steps;  that these contributions have as much practical value as mathematical results;  and that therefore they merit publication as much as the mathematical results our discipline esteems so highly.  

\newpage

\section{Introduction:  Why we should do science as well as math to understand our methods}\label{Intro}

This paper is about how we study statistical methods, using as an example analyses that employ a particular model.  We begin by describing the latter, which leads to the former.

A random regressions model (as it's called in some literature) is a common choice for a situation in which observations are made in clusters and observation $j$ in cluster $i$ has outcome $y_{ij}$ that we would like to model as a linear function of a regressor $x_{ij}$.  Ruppert et al (2003, Section 4.2) give the example of weights measured for nine successive weeks on each of 48 young pigs:  a pig is a cluster, pig $i$'s weight in week $j$ is $y_{ij}$, and the regressor $x_{ij}$ is week number.  In general, $x_{ij}$ can be a vector;  this paper considers scalar $x_{ij}$.  Often it makes sense to let the regression relationship vary between clusters as $y_{ij} = \beta_{0i} + \beta_{1i} x_{ij} + \epsilon_{ij}$, where $\epsilon_{ij}$ is an error term;  in the example, $\beta_{1i}$ is pig $i$'s growth rate per week.  If $(\beta_{0i}, \beta_{1i})$ is of interest and clusters have small sample sizes, it can be advantageous to let clusters ``borrow strength" from each other by using a model like the following.  For $y_{ij}$ measured on a continuous scale, the $\epsilon_{ij}$ are modeled as independently and identically distributed (iid) $N(0,\sigma^2_e)$ random variables and the cluster-specific intercept-slope pairs are modeled as iid draws $(\beta_{0i},\beta_{1i})' \sim N((b_{0},b_{1})',\Sigma)$.  In this paper, the $2 \times 2$ covariance matrix $\Sigma$ is parameterized as
\begin{equation}\label{paramzn}
\Sigma = \left[
\begin{array}{cc}
\sigma^2_c             & \rho \sigma_c \sigma_s \\
\rho \sigma_c \sigma_s & \sigma^2_s
\end{array}
\right],
\end{equation}
\noindent where the subscripts ``c" and ``s" refer to the inter\underline{c}epts $\beta_{0i}$ and \underline{s}lopes $\beta_{1i}$ respectively.  The correlation $\rho$ describes the association, across clusters, of $\beta_{0i}$ and $\beta_{1i}$.

The conventional analysis of a mixed linear model like this begins by maximizing the restricted likelihood (sometimes called the residual likelihood) to estimate $(\sigma^2_e,\sigma^2_c, \sigma^2_s, \rho)$.  The estimates $(\hat{\sigma}^2_e, \hat{\sigma}^2_c, \hat{\sigma}^2_s, \hat{\rho})$ are then taken as given and estimates and tests for, e.g., $(b_{0},b_{1})$ are computed, so $(\hat{\sigma}^2_e, \hat{\sigma}^2_c, \hat{\sigma}^2_s, \hat{\rho})$ is central to the analysis.  A Bayesian analysis adds prior distributions for $(b_{0},b_{1})$ and $(\sigma^2_e,\sigma^2_c, \sigma^2_s, \rho)$ to give posterior distributions and other summaries, but problems afflicting the conventional analysis are still of interest because the restricted likelihood, multiplied by a prior for $(\sigma^2_e, \sigma^2_c, \sigma^2_s, \rho)$, is identical to the marginal posterior for $(\sigma^2_e, \sigma^2_c, \sigma^2_s, \rho)$ assuming a flat (improper) prior for $(b_{0},b_{1})$.  If $(\sigma^2_e, \sigma^2_c, \sigma^2_s, \rho)$ also has a flat prior, the marginal posterior is identical to the restricted likelihood.

The restricted-likelihood maximizing $(\hat{\sigma}^2_e, \hat{\sigma}^2_c, \hat{\sigma}^2_s, \hat{\rho})$ can be on the boundary of legal values, i.e., $\hat{\rho} = -1$ or $+1$, $\hat{\sigma}^2_c = 0$, or $\hat{\sigma}^2_s = 0$;  based on doing or supervising about 40 such analyses, it often is.  An informal observation is that when such inconvenient results occur, the restricted likelihood is often quite flat, so that the data provide little information about $(\sigma^2_e,\sigma^2_c, \sigma^2_s, \rho)$.  This implies that in a Bayesian analysis, the posterior differs little from the prior.  

Is it a problem that $\hat{\rho} = -1$ or $+1$ fairly readily in practice?  Yes:  In the pig-weight example, if our estimate tells us the slope and intercept for each piglet are perfectly anti-correlated in the population of pigs, this is obviously false.  The defective estimate may be a symptom of a problem with the conventional analysis or the model or experimental design but it is nonetheless substantive nonsense and that is a problem.    

Such estimates are also a problem because when they occur, standard software gives useless or misleading information.  It does so because as a discipline we have practically no knowledge about why such estimates occur:  there is, apparently, no literature on when $\hat{\rho} = -1$ or $+1$, and very little on estimates at boundaries more generally\footnote{The entirety, it seems, of literature on zero variance estimates is discussed below.}.  But perhaps this hole in our theory is not really a problem:  when an inconvenient estimate occurs, one can, for example, examine the profiled log-likelihood or do a Bayesian analysis.  Either alternative, however, leaves us floating on the same sea of ignorance.  If we consider how completely single-error-term linear models are understood --- they have produced no surprises since the 1980s\footnote{S. Weisberg, personal communication, 2016.} --- it is clear by contrast that our discipline has a shortage of understanding about how contemporary methods, even simple ones like random regressions, map data into inferential summaries.  We need more understanding, not just more convenient software.

Why do we, as a discipline, have so little understanding of the methods we have created and promote?  Our primary tool for gaining understanding is mathematics, which has obvious appeal:  most of us trained in math and there is no better form of information than a theorem that establishes a useful fact about a method.  But the preceding sentence imposes a heavy burden:  it must be possible to prove a theorem and facts established by the theorem must be useful.  We find finite-sample facts indispensible because real datasets have finite samples and asymptotic theorems never tell us how to apply their conclusions to finite samples.  But finite-sample theorems about contemporary methods are rare;  it seems inescapable that they are at least extremely difficult, given their popularity in earlier eras.

This paper considers a complementary tool for opening our black-box methods, modeled explicitly on the approach molecular biologists use to open Nature's black boxes.  

Before doing so, it seems necessary to address our discipline's prejudice in favor of mathematics and against such empirical approaches.  Among statisticians, a common response to our suggestion of the molecular-biology model of inquiry is to propose new formulations of the random-regressions model, i.e., to try to turn it into a solvable math problem.  While this {\it might} be productive some day, random regressions is a simple model by today's standard and one can only speculate about whether such an effort will, in fact, produce anything useful even for such a simple model.  

Unfortunately, we statisticians either do not perform or do not publish purely empirical studies.  (If you doubt this, try to find empirical studies of the accuracy of standard approximations.  Student literature reviews found 0 and 1 publications for logistic and Cox regression respectively;  Clifton 1997 and Huppler Hullsiek 1996.)  The strength of this preference seems odd given that our discipline {\it exists} to help others establish facts in situations in which theorems are impossible.  The crux seems to be an implicit view that an empirical approach to studying statistical methods lacks the certainty or intellectual heft to justify publication alongside theorems.  Perhaps this is not surprising:  In the statistical theory used to train us, hypotheses and measurement methods simply exist, when in fact creating them is a real accomplishment (e.g., Kary Mullis's 1993 Nobel prize in chemistry for making polymerase chain reaction practical).  Also, although many, perhaps most, statisticians spend their careers collaborating with scientists, unlike them we have not developed widely-accepted ways of generalizing empirical findings from the specific cases included in experiments.  (It was not necessary, for example, to examine every kind of organism to conclude that the genetic code was the same in all organisms.)  Our dismissive attitude toward empirical studies is also understandable given that in our literature, simulation experiments are rarely more than obligatory but often perfunctory exercises in measuring operating characteristics of methods too complex to permit exact theory, when in fact simulation experiments can be used to test a great variety of hypotheses, as shown below.  

As a matter of strategy, is our discipline or indeed an individual researcher better off doing something relatively simple (the molecular-biological approach) and learning something quickly, or betting on the ability to produce useful facts in the long run with mathematics?  A reasonable strategy would, it would seem, do some of each.  The present paper suggests that we can learn about contemporary black-box statistical methods by mimicking molecular biology, demonstrates that approach, and argues that it makes contributions useful enough to compete with theorems for space in our journals.  Broadly, our colleagues in molecular biology proceed by the following steps:
\begin{itemize}
\item Capture the phenomenon of interest in a simple model system, {\it not} a statistical model but rather ``a usually miniature representation of something" [Merriam-Webster's online dictionary], e.g., an animal or cell-culture model.
\item Hypothesize about the phenomenon of interest in terms of the model system.
\item Do experiments with the model system to test those hypotheses.
\item Iterate, revising the model system and structure of hypotheses as needed.
\item Test the revised hypotheses in a more realistic {\it in vivo} system.
\end{itemize}

\noindent This paper demonstrates this approach by using it to understand why maximizing the restricted likelihood for a random-regressions model often gives boundary-value estimates.  

We are certainly not the first to suggest studying statistical methods empirically.  For example, Larntz (1978) did extensive simulation experiments that established, among many other things, the excessive conservatism of the usual rule of thumb for deciding whether to use the chi-squared approximation for Pearson's chi-squared test.  Larntz's student John Adams (Adams 1990a, 1990b) did massive simulation experiments involving response-surface methods, optimal design, and split-plot designs to produce a great variety of information about linear regression methods, e.g., the effect of variable selection, outlier rejection, and the Box-Cox method on the null distribution of the regression F-statistic.  These studies were carefully designed exercises in measuring operating characteristics;  simulation experiments testing explicit hypotheses are harder to find.  Schapire (2013, 2015) described a series of hypothesis-driven simulation experiments showing, among other things, that no available theory (e.g., Friedman et al 2000) explains why AdaBoost works as well as it does and fails when it does.  (Schapire 2015, a talk in a memorial session for Leo Breiman, emphasized the hypotheses more than does Schapire 2013.  Dr.~Schapire avers [e-mail, 26 August 2015] that Prof.~Breiman ``very much advocated \dots an experimental approach to machine learning/statistics" though we have not found a suitable citation.)  

We emphatically do not claim to have the last word on how to study statistical methods empirically, or to present an algorithm for conducting such studies, an idea that has long since been discredited.  (Feyerabend 1993 is just one example.)  The results of an empirical study are, of course, not as iron-clad as a theorem drawing the same conclusions but the odds against producing such theorems are discouraging.  This suggests that an orchard of low-hanging fruit awaits if our discipline re-directs some energy away from asymptotic theorems and toward this simple, productive approach.

Sections~\ref{modelSystem} through~\ref{inVivo} describe and demonstrate the steps of the molecular-biology approach.  The statistical-methods question is:  In a random-regressions fit, which features of the design or data-generating process influence the chance that the restricted likelihood is maximized at $\hat{\rho} = \pm 1$?  Section~\ref{moreDisc} discusses the results and their implications for how we study our methods.  Each section discusses its step's intellectual content;  although the primary contribution of work in this approach will be the utility of the results, sometimes an empirical study introduces a design of broader value that is itself a contribution.  Details are in the Supplement or are omitted;  all are suitable for student exercises.

\section{The model system}\label{modelSystem}

In biology, model systems include cell cultures and living organisms of complexity ranging from bacteria to zebrafish to rodents to primates.  For our example problem, the model system is a simple random-regressions model specified as follows.

For clusters indexed by $i$, the data $y_{ij}$ are presumed to arise as 
\begin{equation}\label{model1}
y_{ij} = \beta_{0i} + \beta_{1i} x_{ij} + \epsilon_{ij}, \quad i = 1, \dots, N, \quad j = 1, \dots, s,
\end{equation}
\noindent where the $\epsilon_{ij}$ are iid $N(0,\sigma^2_e)$ independently of the $(\beta_{0i},\beta_{1i})'$, which are iid $N((b_{0},b_{1})',\Sigma)$, with $\Sigma$ parameterized as in (\ref{paramzn}).  The cluster size is $s = 2m+1$ for $m$ a positive integer, so $s$ is odd.  Given $s$, the regressor $x_{ij}$ is stacked in a vector $\bh$ taking the value 
\begin{equation}\label{xValues}
\bh = (-1, -(m-1)/m, \dots, 0, \dots, (m-1)/m, 1)'
\end{equation}
\noindent so the range of the $x_{ij}$ does not change with $s$.  (This prevents some uninteresting artifacts.)  

To put this in the usual mixed linear model notation $\by = \bX \bbeta + \bZ \bu + \beps$ with $cov(\bu) = \bG$ and $cov(\beps) = \bR$, let $\bH$ be the $s \times 2$ design matrix within a cluster, with orthogonal columns $\bone_s$, an $s$-vector of 1's, and $\bh$ as in (\ref{xValues}).  We have $N$ clusters and $n = Ns$ observations, and
\begin{equation}\label{specification}
\begin{array}{ll}
\bX = \bone_N \otimes \bH & \mbox{ is } n \times 2 \nonumber \\
\bbeta = (b_{0},b_{1})' & \mbox{ is } 2 \times 1 \nonumber \\
\bZ = \bI_N \otimes \bH & \mbox{ is } n \times 2N \nonumber \\
\bu = (u_{10},u_{11} | u_{20},u_{21} | \dots | u_{N0},u_{N1} )' & \mbox{ is } 2N \times 1 \nonumber \\
\bG = \mbox{BlockDiag}(\Sigma)  & \mbox{ is } 2N \times 2N, \mbox{ and }  \nonumber \\
\bR = \sigma^2_e \bI_n, 
\end{array}
\end{equation}
\noindent where the observations in $\by$ are sorted first by cluster $i$ and then by $j$ within cluster, $\otimes$ is the Kronecker product, defined as $\bA \otimes \bB = (a_{ij}\bB)$, ``$|$" indicates matrix partitioning, and $\bI_d$ is the $d$-dimensional identity matrix.  In this notation, $\beta_{0i} = b_0 + u_{i0}$ and $\beta_{1i} = b_1 + u_{i1}$.  

These features of the model system can be varied:  the number of clusters $N$, the within-cluster sample size $s$, and all unknowns.  Nature presents us with $(\sigma^2_e,\sigma^2_c, \sigma^2_s, \rho)$;  sometimes, at least, we can hope to manage the chance of $\hat{\rho} = \pm1$ by choosing the sample sizes $N$ and $s$.  The simulation experiments below set $(b_{0},b_{1})' = (0,0)'$ without loss of generality because this mean structure is removed from the data in computing the restricted likelihood.

What is this step's intellectual content?  The achievement in choosing a model system (or a ``simplest interesting case") is to make it as simple as possible while still allowing hypotheses of interest to be stated in the model's terms and tested.  In biology, the payoffs of simplification are ability to control inputs to and measure outputs of the model system, and to isolate the causal effects of factors manipulated in experiments.  Simplification has a cost, the need to hedge on interpretation;  reviews of grant proposals and journal manuscripts generally involve implicit negotiation about the limits of the model system.  For studying statistical methods, the payoffs of simplification are ability to do more explicit derivations and simpler, faster computing, which permit hypotheses to be formulated and tested economically in designed experiments.  The cost is the risk of omitting an important feature of the problem, though the final step (``test the revised hypothesis \dots `{\it in vivo}'") provides some protection against this risk.  The model system above is simplified by having a single regressor and forcing all clusters to have the same design matrix, with orthogonal columns.  These simplifications and others introduced below permit fairly explicit derivations while retaining the ability to manipulate $N$, $s$, and $(\sigma^2_e,\sigma^2_c, \sigma^2_s, \rho)$.  As in biology, the key questions are whether the model system reproduces the phenomena of interest and whether it behaves like the unsimplified system.  We will see below that for the present model system the answers are, respectively, ``yes" and ``as far as we can tell".  

For this model, the restricted likelihood is straightforward, so its derivation is omitted.  Define $\bL$ to be $s \times (s-2)$ with orthonormal columns satisfying $\bL'\bone_s = \bL'\bh = 0$; $\bL$ is used to project onto the residual space within each cluster.  Define $\bW$ to be $N \times (N-1)$ with orthonormal columns such that $\bW'\bone_N = \bze$;  $\bW$ is used to project onto the between-cluster residual space and its columns are contrasts in cluster-specific quantities.  The log restricted likelihood is then
\begin{eqnarray}\label{logRL}
-\frac{N(s-2)}{2}\log \sigma^2_e &-&\frac{1}{2\sigma^2_e} RSS \label{errPart}\\ 
- \frac{N-1}{2}\log \det(\bF) &-&\frac{1}{2} \sum_{k=1}^{N-1} \bt_k' \bF^{-1} \bt_k, \label{GPart} \\
\mbox{where } RSS &=& \by' (\bI_N \otimes \bL \bL') \by \label{RSS}\\
\mbox {and } \bF &=& \bD \Sigma \bD + \sigma^2_e \bI_2 \quad \mbox{for} \quad \bD =\mbox{diag}(s,q)^{1/2}, \nonumber
\end{eqnarray}
\noindent where $q = (2m^2+3m+1)/3m$ and the $2 \times 1$ vector $\bt_k = (\bW_k' \otimes [\bD^{-1}\bH'])\by$ for $\bW_k$ the $k^{th}$ column of $\bW$.  Line (\ref{errPart}) is a function only of $\sigma^2_e$ and $\by$; the quadratic form $RSS$ in (\ref{errPart}) is the residual sum of squares from the unshrunk regressions within clusters, aggregated across clusters.  The unknowns $(\sigma^2_c, \sigma^2_s, \rho)$ appear only in (\ref{GPart}), in which each $\bt_k$ is a contrast in the unshrunk estimates of cluster-specific $(\beta_{0i},\beta_{1i})$.  The sum of quadratic forms in (\ref{GPart}) is the additional residual sum of squares arising from shrinkage for given $(\sigma^2_e, \sigma^2_c, \sigma^2_s, \rho)$.

This form of the log restricted likelihood is about as simple as possible but it is still too complicated to allow intuition:  $\bF$ is a function of all of $\sigma^2_e, \sigma^2_c, \sigma^2_s$, and $\rho$ and the data enter both (\ref{errPart}) and (\ref{GPart}) in complicated ways.  This affects the next step in the biological model of inquiry, generating hypotheses, to which we now turn.  

\section{Generating hypotheses}\label{genHyp}

A necessary condition for the log restricted likelihood (henceforth log RL) or profiled log RL to be maximized at $\hat{\rho} = -1$ is that its derivative with respect to $\rho$, evaluated at $\rho = -1$, is negative.  This condition motivates deriving a {\it predictor} of $\hat{\rho} = -1$ to use in generating hypotheses, as follows:  simplify the log RL, profile out one unknown, evaluate the derivative of the profiled log RL with respect to $\rho$ at $\rho = -1$, and finish with more simplifications.  We begin to deviate from a mathematical approach here by creating an approximation not to replace the original expression but as an instrument for generating hypotheses.  This grants us a certain freedom:  the predictor's utility lies not in its accuracy as an approximation but rather in the frequency with which simulation experiments support the hypotheses it helps us generate, and this utility is exhausted when the hypotheses are tested.

The first simplification is to let $\sigma^2_c = \sigma^2_s \equiv \sigma^2_r$.  Real problems can be made close to this by scaling $x_{ij}$, so this seems like a small sacrifice in generality.  The log RL now has three unknowns, $\sigma^2_r$, $\sigma^2_e$, and $\rho$;  we reduce that to two by profiling out $\sigma^2_r$.   Define $r = \sigma^2_e / \sigma^2_r$ so that $\sigma^2_e = r \sigma^2_r$.  It is easy to show that the maximizing value of $\sigma^2_r$ given $(r, \rho)$ is 
\begin{equation}
\hat{\sigma}^2_r(r, \rho) = (Ns-2)^{-1}\{RSS/r + \sum_{k=1}^{N-1} \bt_k'(\sigma^2_r\bF^{-1}) \bt_k\};
\end{equation}
\noindent note that $\sigma^2_r\bF^{-1}$ is a notational convenience and is not in fact a function of $\sigma^2_r$.  Ignoring unimportant constants, the profiled log RL is then
\begin{eqnarray}
&& -\frac{N(s-2)}{2}\log r  -\frac{N-1}{2}\log\{ (s+r)(q+r) - sq\rho^2 \} \label{logProfRL}\\
&& -\frac{Ns-2}{2}\log\{RSS/r + \sum_{k=1}^{N-1} \bt_k'(\sigma^2_r\bF^{-1}) \bt_k\}. \nonumber
\end{eqnarray}

The derivative of (\ref{logProfRL}) with respect to $\rho$ at $\rho = -1$ is messy and permits little intuition.  Therefore, we simplified the derivative by replacing functions of the data with their expected values under the model.  Specifically, we replaced $r/RSS$ by $(N(s-2)\sigma^2_r)^{-1}$ and, defining $\bt_k = (t_{kc},t_{ks})'$, we replaced $t_{kc}^2$ by $\sigma^2_r(s+r)$, $t_{ks}^2$ by $\sigma^2_r(q+r)$, and $t_{kc}t_{ks}$ by $\sqrt{sq}\sigma^2_r\rho$, yielding this predictor of when $\hat{\rho} = -1$:
\begin{eqnarray}\label{predictor}
\left(\frac{Ns - N - 1}{ (1+r/s)(1+r/q) }\right)
\left[
1 - \left(\frac{Ns-2}{Ns-N-1}\right) \frac{1-\frac{N-1}{N(s-2)}\rho}{1 + \frac{2(N-1)}{N(s-2)} 
\frac{(1+r/s)(1+r/q) + \rho}{(1+r/s)(1+r/q) -1}      }
\right],
\end{eqnarray}
\noindent a function of $N$, $s$, $\rho$, and $r$ but not $\sigma^2_r$.  This can be understood as a $0^{th}$ order Taylor expansion around the expected values of functions of the data.  (It is easy to derive a very similar predictor for $\hat{\rho} = +1$, which is given in the Supplement and used below.)

We seek to understand when the derivative of the log RL (\ref{logRL}) with respect to $\rho$ at $\rho = -1$ is negative; we derived the predictor (\ref{predictor}) to generate hypotheses about when that derivative is negative.  However, it's easy to show that the predictor is positive for all finite $N$ and $s$, all $\rho \in (-1,1)$, and all finite positive $r$.  (The proof is in the Supplement.)  Thus (\ref{predictor}) fails as an {\it approximation} to the derivative of the profiled log RL, but as we'll see in the simulation experiments, it does well at predicting when $\hat{\rho} = -1$:  as  (\ref{predictor}) becomes smaller (closer to zero), $\hat{\rho}$ is more likely to be $-1$ and as  (\ref{predictor}) becomes larger, $\hat{\rho}$ is less likely to be $-1$.  Thus it is accurate to describe (\ref{predictor}) as a predictor of the event $\hat{\rho} = -1$.  

We now exercise the predictor (\ref{predictor}) as a function of $N$, $s$, $\rho$, and $r$ to generate hypotheses about what promotes or suppresses $\hat{\rho} = \pm1$.  The following facts about the predictor of $\hat{\rho} = -1$ are easy to prove (proofs are in the Supplement):
\begin{itemize}
\item Given $N$, $s$, and $\rho$, as $r$ increases --- i.e., as the error variance $\sigma^2_e$ increases relative to the random-effect variance $\sigma^2_r$ --- the predictor goes to zero.
\item Given $\rho$ and $r$, as either $N$ or $s$ increases, the predictor increases.
\item Given $N$, $s$, and $r$, as $\rho$ goes to $-1$, the predictor goes to zero.  
\end{itemize}
\noindent Combined with the hypothesis that a small predictor value implies a high chance that $\hat{\rho} = -1$ and a large predictor value implies a small chance that $\hat{\rho} = -1$, these facts give three hypotheses about $\hat{\rho}$'s behavior.  If these hypotheses are correct --- and the simulation experiments support them --- then $\hat{\rho} = -1$ becomes more likely as the error variance increases relative to the intercept and slope variances and less likely as either sample size increases.  This implies that $\hat{\rho} = \pm1$ is mainly a consequence of poor resolution in the study design, with ``resolution" used in the same sense as the resolution of a measuring device or video monitor, i.e., error variation is large and is not suppressed by sample size.  (The tiny literature about zero variance estimates is consistent with this.  For the balanced one-way random effects model, an estimate of zero for the between-groups variance arises from poor resolution;  see Hill 1965, Section 3B.  Hodges 2014, Chapter 18 extends Hill 1965 and shows an example suggesting that the same is true for mixed-effects analysis of variance more generally.)  

To develop quantitative hypotheses about how $N$, $s$, $\rho$, and $r$ affect the chance of $\hat{\rho} = -1$, we drew 1000 sets of $(N, s, \rho, r)$ by making independent draws of each quantity and computed the common (base 10) log of the predictor for each such set.  Each quantity was drawn iid, $N$ from $\{50, 150, 250, \dots, 1050\}$, $s$ from $\{5, 15, 25, \dots, 105\}$, $\rho$ from $\{-0.9, -0.8, \dots, -0.1\}$, and $\log_{10}r$ from $\{-2, -1.6, -1.2, \dots, 2\}$.  The resulting design in $(N, s, \rho, r)$ was roughly balanced.  (Real datasets we have analyzed as random regressions had cluster counts and sizes at the low ends of these ranges of $N$ and $s$.)  Analyzing these $\log_{10}$ predictor values using ANOVA with factors $N$, $s$, $\rho$, and $r$, the main effects for $r$, $s$, $N$, and $\rho$ had mean squares 172.5, 23.0, 5.3, and 2.0 respectively;  the two-way interactions $\rho$-by-$r$ and $s$-by-$r$ had mean squares 0.08 and 0.01 respectively;  and the other four two-way interactions and the combined three- and four-way interactions each had mean squares less than $2 \times 10^{-5}$.  Thus the main effects dominate the behavior of $\log_{10}$ predictor.  

\begin{figure}
   \centering
   \begin{subfigure}[b]{0.45\textwidth}
       \includegraphics[clip=true, trim= 140 65 170 50, width=\textwidth]{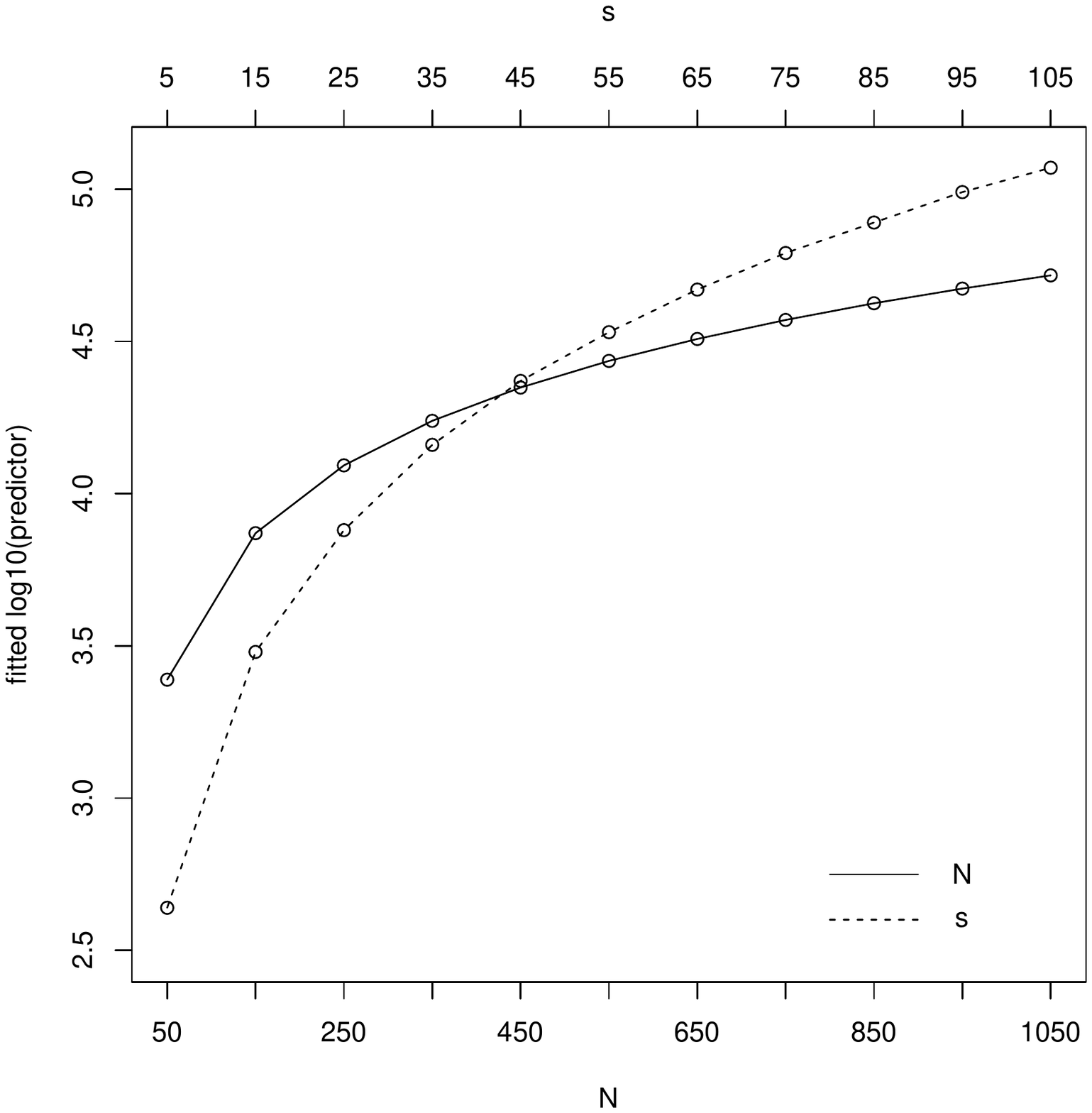}
       \caption{}
       \label{schnell}
   \end{subfigure}
   \begin{subfigure}[b]{0.45\textwidth}
       \includegraphics[clip=true, trim= 140 65 170 50, width=\textwidth]{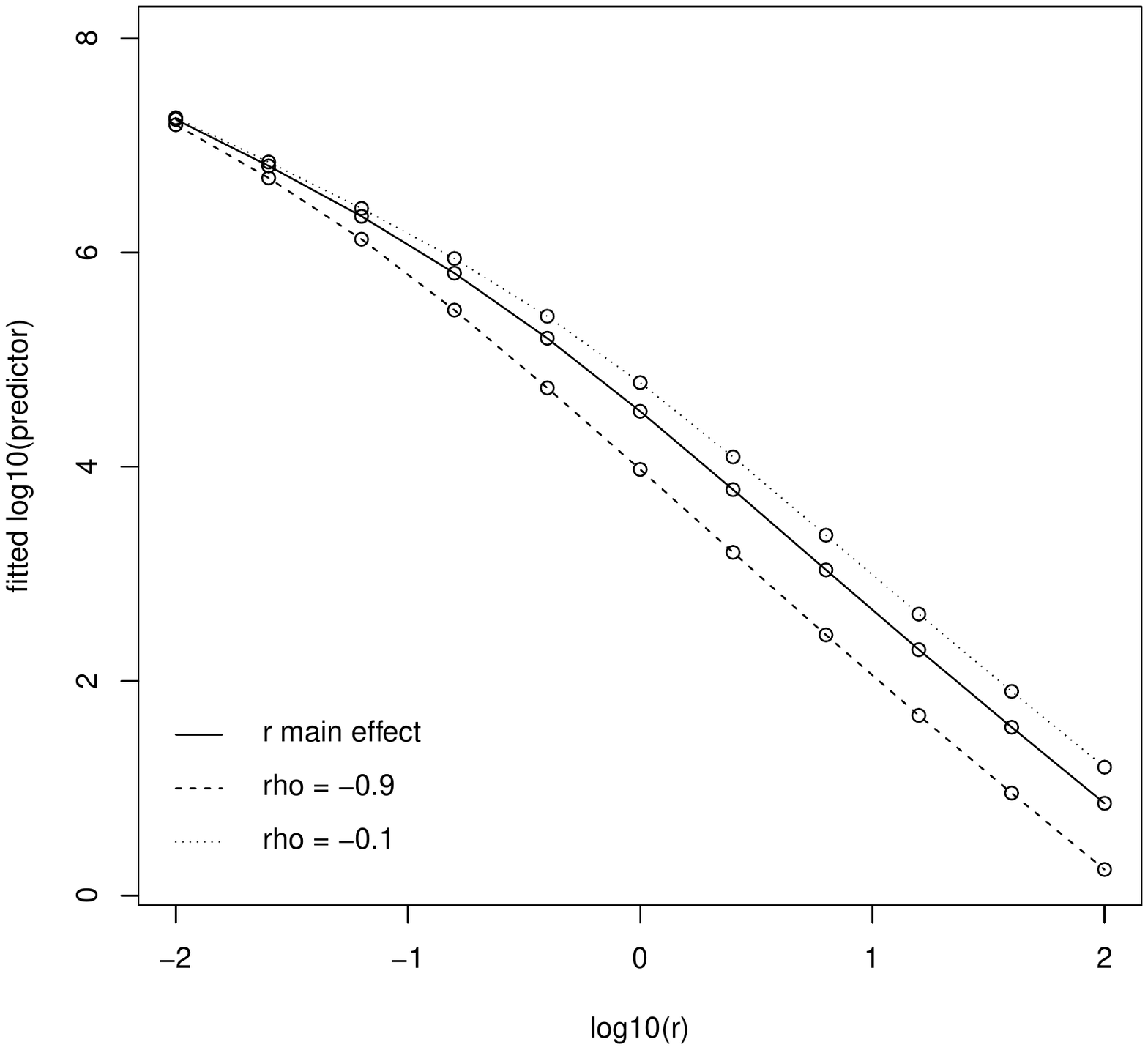}
       \caption{}
       \label{rRho}
   \end{subfigure}
   \caption{Exercising the predictor.  Panel (a):  Solid line:  estimated main effect of $N$, referring to the axis at the plot's bottom.  Dashed line: estimated main effect of $s$, referring to the axis at the plot's top.  Panel (b):  Solid line:  estimated main effect of $r$.  Other lines:  Estimated effect of $r$ for $\rho$ = $-0.9$ (dashed) and $-0.1$ (dotted).}
\label{genHypRes}
\end{figure}

Figure~\ref{schnell} shows the effects of $N$ and $s$ marginal to (i.e., averaging over) the other factors;  its vertical axis is SAS's least-squares means in an analysis treating $N$, $s$, $\rho$, and $r$ as categorical factors and including the main effects and all six two-way interactions.  For a given proportionate increase in $N$ or $s$, the predictor is increased more by increasing $s$ than $N$.  Because $\log_{10}$ predictor increases at a diminishing rate as $N$ or $s$ increases, one might suspect that $N$'s effect is smaller because larger $N$ values were considered, but this is not so:  increasing $N$ from 50 to 150 increases $\log_{10}$ predictor by 0.48, while increasing $s$ from 55 to only 105 increases $\log_{10}$ predictor by 0.54.  

Figure~\ref{rRho} shows the main effect of $r$ and the $r$-by-$\rho$ interaction.  The effect of increasing $\log_{10}r$ by 0.4 (multiplying $r$ by about 2.5) grows as $r$ becomes larger.  The difference between the lines for $\rho = -0.1$ and $-0.9$ depends on $r$:  If $r$ is small (the design has good resolution), $\rho$ has little effect on the predictor, while if $r$ is large (the design has poor resolution) $\rho$ has some effect.  However, even for $r = 100$, the effect of this change in $\rho$, about a 1-log change in $\log_{10}$ predictor, is less than the effect of increasing $r$ from 16 to 100.  Thus, if $r$ is made large enough, it can overwhelm any benefit produced by Nature's choice of, say, $\rho=0$.  

Here are some further hypotheses arising from this exercise.
\begin{itemize}
\item The effect of an increase or decrease in $r$ can be countered by increasing or decreasing (respectively) $N$ or $s$.  Multiplying or dividing $r$ by about 2.5 is countered by multiplying or dividing (respectively) $N$ by about 5 or $s$ by about 3.  If confirmed (and it is), this hypothesis has a design implication:  for a given increase in total sample size, increasing $s$ causes a greater reduction in the chance that $\hat{\rho} = -1$ than increasing $N$.  
\item Changes in $\rho$ induce smaller changes in the chance that $\hat{\rho}=-1$ than do changes in $r$, $N$, or $s$;  some changes in $r$ have effects so large that no change in $\rho$ can counter them.  In particular, if $r$ is large enough then for any $\rho$, $\hat{\rho}$ is very likely to be $-1$.  
\end{itemize}

As for this step's intellectual content, hypothesis generation is one of the central creative activities of scientific work and a key difference between competent and brilliant scientists is that the latter pose more fruitful and penetrating hypotheses.  The ability to produce deep, powerful hypotheses depends on insight and creative manipulation of the method under study.  Molecular biologists can now generate and test hypotheses by manipulating their objects of study, e.g., by creating gene-knockout organisms;  we can generate hypotheses by manipulating our objects of study, which are combinations of equations and algorithms, using approximations as above.  

The mathematical approach to studying statistical methods does include hypotheses;  they are called unproven conjectures and rarely see the light of day unless they are proven, while disproofs of scientific hypotheses are routinely published.  If empirical study of statistical methods became more common, it might not be appropriate for journal articles to describe the hypothesis-generation step at length, as we have, or perhaps to describe it at all but its importance cannot be denied.  

\section{Testing the hypotheses using simulation experiments}\label{symExpts}

With hypotheses in hand, the next step is to design and execute experiments to test them. 

To derive the predictor, we set $\sigma^2_c = \sigma^2_s$ and in the simulation experiments below, we simulated data by setting $\sigma^2_c = \sigma^2_s$.  The goal, however, is to understand the log LR-maximizing estimates for the unsimplified random-regressions model, in which $\sigma^2_c$ and $\sigma^2_s$ can be different.  Thus, although the hypotheses were stated above in terms of a single variance $\sigma^2_r$ for both the random intercept and random slope, and thus in terms of $r = \sigma^2_e/\sigma^2_r$, and data generation in the experiments below has $\sigma^2_c = \sigma^2_s$, this restriction was {\it not} enforced in fitting the models.

We did preliminary simulations to determine predictor values that are ``on the cusp", i.e., $(N, s, \rho, r)$ having these predictor values give some but not too many bad estimates, where a ``bad" estimate\footnote{We use this word at the risk of offending readers because it is short and, as argued above, appropriate.} is $\hat{\rho} = \pm1$ or $\sigma^2_c=0$ or $\sigma^2_s=0$.  This is the region of $(N, s, \rho, r)$ in which changes in these inputs can affect the fraction of bad estimates, so it is most useful for testing the hypotheses.  Beginning this way mimicks our colleagues in biology:  they select experimental settings so that their model system's response is middling and thus most readily changed by manipulating inputs.

The experiments are in three groups:  one testing whether increasing $r$, the ratio of the error variance to the random-effect variances, produces a large fraction of bad estimates;  one examining the tradeoffs between $r$ on the one hand and $N$, $s$, and $\rho$ on the other hand;  and a final set examining the effect of $\rho$.  We present these three sets of experiments in turn.  

The methods for the experiments were as follows.
\begin{itemize}
\item Datasets were simulated from (\ref{model1}) to (\ref{specification}) with $\bbeta = (0,0)'$ and $\sigma^2_c = \sigma^2_s = 1$ (with one exception to the latter, noted below).  Counts of simulated datasets in each experiment are given with the results.  
\item Analyses were done in R (v.~3.1.2, R Core Team 2014) using the lmer function (lme4 package v.~1.1-7, Bates et al 2014).  In preliminary experiments, lmer always found a local maximum in constrast to various options in the nlme package, which failed sometimes for large $r$.  The variances $\sigma^2_c$ and $\sigma^2_s$ were not forced to be equal in the fit.
\item Results of experiments are summarized by presenting the values of $N$, $s$, $\rho$, and $r$ defining each experimental setting, the predictors for $-1$ and $+1$, the percent of estimates $\hat{\rho} = -1, +1$, and NaN (``not a number" in R;  i.e., $\sigma^2_c=0$ or $\sigma^2_s=0$), and the sum of these three percents, i.e., the percent of bad estimates.  
\end{itemize}

\subsection{Increasing $r$ produces bad estimates}

Table~\ref{bigr} shows results from Experiments A and B, which were done specifically for this hypothesis though other experiments (below) support the same conclusion, especially Experiment G.  We used $N = 500$ and $s = 21$ in these experiments, values quite a bit larger than in any real dataset to which we have fit a random regressions model.  In both experiments, as $r$ increases, the predictor decreases and the percent of simulated datasets yielding bad estimates increases.  Datasets giving $\hat{\rho} =$ NaN had $\hat{\sigma}^2_c = 0$ in all cases examined.  

The difference between Experiments A and B is the value of $\rho$ used in simulating the data.  Some might conjecture that the chance of a bad estimate is minimized by setting $\rho = 0$ while the chance of $\hat{\rho}= -1$ would be reduced by setting $\rho$ close to +1;  Experiments A and B (respectively) use these $\rho$.  In these experiments, setting $\rho$ close to +1 has no effect on the chance of $\hat{\rho}= -1$ for the two largest $r$ values (settings 4 and 5) and only a slight effect for setting 3.  However, setting $\rho$ close to +1 does give a much higher chance of $\hat{\rho}= +1$ in settings 1 and 2, and thus increases the chance of a bad estimate for those settings.

(In generating data for these experiments, $r$ was made large by fixing $\sigma^2_c = \sigma^2_s = 1$ and making $\sigma^2_e$ large but $r$ could also be made large by fixing $\sigma^2_e$ and making $\sigma^2_c$ and $\sigma^2_s$ small.  We re-did Experiments A and B with all settings identical except that $\sigma^2_e$ was fixed at 1 and $\sigma^2_c$ and $\sigma^2_s$ were set to give the desired $r$.  The results, with 400 simulated datasets per setting, were indistinguishable from those in Table~\ref{bigr}.)

Experiment A gives the first hint of an oddity that recurs in later experiments:  With $\rho = 0$ and large $r$, we would expect $\hat{\rho}= +1$ and $\hat{\rho}= -1$ to be about equally likely, but in fact $\hat{\rho}= +1$ is rather more likely.  In each of Experiments A and B, combining settings 3, 4, and 5, the fractions of datasets with $\hat{\rho}= +1$ tests higher than the fraction with $\hat{\rho}= -1$ ($P < 0.001$ in a two-tailed test).  

\begin{table}
\caption{Simulation experiments:  Increasing $r$ increases the chance of a bad estimate.  100 datasets per setting.}\label{bigr}
\begin{center}
\begin{tabular}{c|cccc|cc|rrrr}
\multicolumn{5}{l}{{\bf Experiment A}} & \multicolumn{2}{|c|}{predictor} & \multicolumn{4}{|c}{\% with $\hat{\rho}$} \\ 
setting & $N$   & $s$  & $\rho$ & $r$    & -1     &  +1    & -1 & +1 & NaN & Bad     \\ \hline
 1  & 500 & 21 &   0 & $10^1$ & 3.6e+2 & 3.6e+2 &  0 &  0 &   0 &   0  \\
 2  & 500 & 21 &   0 & $10^2$ & 6.4e+0 & 6.4e+0 & 11 &  7 &   1 &  19  \\
 3  & 500 & 21 &   0 & $10^3$ & 8.0e--2 & 8.0e--2 & 21 & 33 &  30 &  84  \\
 4  & 500 & 21 &   0 & $10^4$ & 8.2e--4 & 8.2e--4 & 19 & 29 &  40 &  88  \\
 5  & 500 & 21 &   0 & $10^5$ & 8.2e--6 & 8.2e--6 & 15 & 36 &  32 &  83  \\ \hline
\multicolumn{11}{c}{} \\ 
\multicolumn{5}{l}{{\bf Experiment B}} & \multicolumn{2}{|c|}{predictor} & \multicolumn{4}{|c}{\% with $\hat{\rho}$} \\ 
setting & $N$   & $s$  & $\rho$ & $r$    & -1     &  +1    & -1 & +1 & NaN & Bad     \\ \hline
  1 & 500 & 21 & 0.95 & $10^1$ & 6.8e+2 & 1.5e+1 &  0 & 25 &  0 & 25  \\
  2 & 500 & 21 & 0.95 & $10^2$ & 1.3e+1 & 2.6e--1 &  0 & 54 &  3 & 57  \\
  3 & 500 & 21 & 0.95 & $10^3$ & 1.5e--1 & 3.1e--3 & 17 & 31 & 29 & 77  \\
  4 & 500 & 21 & 0.95 & $10^4$ & 1.6e--3 & 3.2e--5 & 22 & 37 & 31 & 90  \\
  5 & 500 & 21 & 0.95 & $10^5$ & 1.6e--5 & 3.2e--7 & 21 & 36 & 29 & 86  \\ \hline
\end{tabular}
\end{center}
\end{table}

\subsection{Trading off $r$ against $N$, $s$, and $\rho$}

Table~\ref{TradeTable} shows Experiments C through F, which have the same structure:  Setting 1 is a base case chosen to give some bad estimates;  in setting 2, $r$ is changed;  and settings 3, 4, and 5 attempt to reverse the effect of setting 2's change in $r$ by changing only $N$, $s$, or $\rho$ respectively.  In Experiments C, E, and F, setting 2 had a larger $r$ than in setting 1, giving more bad estimates;  in Experiment D, setting 2 had a smaller $r$ than in setting 1, giving fewer bad estimates.  For setting 2, $r$ was either larger or smaller by a factor of about 2.5 ($10^{0.4}$) compared to setting 1, and $N$, $s$, and $\rho$ in settings 3, 4, and 5 were chosen to make the predictor of $\hat{\rho}= -1$ as similar as possible to its value in setting 1.  (Sometimes setting 4's predictor value could not hit the target because $s$ must be odd and at least 3.)  

\begin{table}
\caption{Changing $N$, $s$, and $\rho$ to counter changes in $r$.  400 datasets per setting except Experiment D, which had 600.}\label{TradeTable}
\begin{center}
\begin{tabular}{c|cccc|cc|rrrr}
\multicolumn{5}{l}{{\bf Experiment C}} & \multicolumn{2}{|c|}{predictor} & \multicolumn{4}{|c}{\% with $\hat{\rho}$} \\ 
setting & $N$   & $s$  & $\rho$ & $r$    & -1     &  +1    & -1 & +1 & NaN & Bad     \\ \hline
1  &  100&   9&  -0.8&   6.3&     8.2&    67&    18&     0&     0&    18 \\
2  &  100&   9&  -0.8&   15.8&     1.7&    15&    39&     0&     2&    41 \\
3  &  500&   9&  -0.8&   15.8&     8.4&    74&    17&     0&     0&    17 \\
4  &  100&  25&  -0.8&   15.8&     8.7&    76&    20&     0&     0&    20 \\
5  &  100&   9&   0.0&   15.8&     8.2&   8.2&     8&     6&    17&    30 \\ \hline
\multicolumn{11}{c}{} \\ 
\multicolumn{5}{l}{{\bf Experiment D}}  & \multicolumn{2}{|c|}{predictor} & \multicolumn{4}{|c}{\% with $\hat{\rho}$} \\ 
setting & $N$   & $s$  & $\rho$ & $r$    & -1     &  +1    & -1 & +1 & NaN & Bad     \\ \hline
  1 & 100 & 9 & -0.80 & 15.8  & 1.66 & 14.60    & 42    & 1    & 1    & 44 \\
  2 & 100 & 9 & -0.80 &  6.3  & 8.23 & 67.47    & 19    & 0    & 0    & 19 \\
  3 &  21 & 9 & -0.80 &  6.3  & 1.67 & 13.69    & 48    & 1    & 1    & 50 \\
  4 & 100 & 3 & -0.80 &  6.3  & 1.83 & 15.14    & 42    & 0    & 0    & 42 \\
  5 & 100 & 9 & -0.96 &  6.3  & 1.66 & 72.82    & 47    & 0    & 0    & 47 \\ \hline
\multicolumn{11}{c}{} \\ 
\multicolumn{5}{l}{{\bf Experiment E}}  & \multicolumn{2}{|c|}{predictor} & \multicolumn{4}{|c}{\% with $\hat{\rho}$} \\ 
setting & $N$   & $s$  & $\rho$ & $r$    & -1     &  +1    & -1 & +1 & NaN & Bad     \\ \hline
  1 &  20  &  25 & -0.8  &   6 &   9.88 &  79.92 &  23  &   0  &    0 &   23 \\
  2 &  20  &  25 & -0.8  &  15 &   1.85 &  16.02 &  42  &   2  &    2 &   45 \\
  3 & 104  &  25 & -0.8  &  15 &  10.00 &  86.64 &  15  &   0  &    0 &   15 \\
  4 &  20  &  63 & -0.8  &  15 &   9.66 &  83.30 &  23  &   0  &    2 &   25 \\
  5 &  20  &  25 &  0.0  &  15 &   9.06 &   9.06 &   9  &  10  &    4 &   23 \\ \hline
\multicolumn{11}{c}{} \\ 
\multicolumn{5}{l}{{\bf Experiment F}}  & \multicolumn{2}{|c|}{predictor} & \multicolumn{4}{|c}{\% with $\hat{\rho}$} \\ 
setting & $N$   & $s$  & $\rho$ & $r$    & -1     &  +1    & -1 & +1 & NaN & Bad     \\ \hline
  1 & 1000 & 3 & -0.8  &   9 &  10.05 & 86.15  &  24   &  0   &   0  &   24 \\
  2 & 1000 & 3 & -0.8  &  23 &   1.88 & 16.77  &  38   &  0   &   0  &   38 \\
  3 & 5350 & 3 & -0.8  &  23 &  10.08 & 89.81  &  21   &  0   &   0  &   21 \\
  4 & 1000 & 9 & -0.8  &  23 &   8.78 & 77.92  &  19   &  0   &   2  &   20 \\
  5 & 1000 & 3 &  0.0  &  23 &   9.36 &  9.36  &  10   &  6   &   0  &   16 \\ \hline
\end{tabular}
\end{center}
\end{table}

Experiments C and D used $N$ and $s$ that seem moderate though they are larger than in any dataset we've analyzed.  Experiment E has small $N$ and large $s$;  Experiment F has large $N$ and small $s$.  We chose $\rho = -0.8$ for setting 1 in all experiments because in applications, $\rho < 0$ is usually more plausible than $\rho> 0$:  when the average slope in $x_{ij}$ is positive, clusters with low intercepts have more room to increase (and thus larger slopes) than do clusters with high intercepts, and analogously when the average slope in $x_{ij}$ is negative.

The chosen $N$ and $s$ counter the change in $r$ more or less as predicted.  In these four experiments, a test comparing setting 3 vs.~setting 1 (i.e., does the change in $N$ counter the change in $r$?) gives a large (non-significant) P-value except for Experiment E, where the increase in $N$ compensates more than predicted.  Similarly, a test comparing setting 4 vs.~setting 1 (i.e., does the change in $s$ counter the change in $r$?) gives a large (non-significant) P-value in all four experiments.  The proportional changes in $s$ that counter an experiment's change in $r$ were 2.8, 3, 2.5, and 3 in Experiments C, D, E, and F respectively;  the corresponding proportional changes in $N$ were 5, 4.8, 5.2, and 5.4.  

A bigger surprise is the effect of changes in $\rho$, examined further in the following subsection.

\subsection{The effect of $\rho$}

The predictor predicted that a large change in $\rho$, from $-0.8$ to 0 in Experiments C, E, and F, will counter an increase in $r$.  But this change in $\rho$ did not simply reduce the chance that $\hat{\rho}= -1$, though it did that;  it also increased the chance that $\hat{\rho}= +1$ or NaN ($\hat{\sigma}^2_c = 0$ or $\hat{\sigma}^2_s = 0$).  In Experiments C, E, and F, setting 5 with $\rho$ increased from $-0.8$ to 0 produced fewer datasets with $\hat{\rho} = -1$ than settings 3 and 4 but this change in $\rho$ did not predictably counter changes in the chance of {\it any} kind of bad estimate:  sometimes it overcompensated and sometimes it undercompensated.  

Experiment G, summarized in Table~\ref{bigrNrho}, further explores the effect of $\rho$.  All settings have the same fairly large $N$ and $s$.  Each block of 5 settings has one value of $r$ and includes $\rho$ ranging from $-0.95$ to $+0.95$.  For the relatively small $r = 53$, $\hat{\rho}$ behaves as one might expect:  when $\rho = -0.95$, 45\% of the datasets have $\hat{\rho} = -1$, when $\rho = +0.95$, 47\% of the datasets have $\hat{\rho} = +1$, and intermediate $\rho$ give intermediate results.  However, as $r$ increases this tidy pattern dissolves so that when $r$ = 3,000 or 100,000, the true $\rho$ does not matter:  $\hat{\rho} = -1$ is about equally likely for all $\rho$, as are $\hat{\rho} = +1$ and NaN.  

\begin{table}
\caption{If $r$ is big enough, $\rho$ doesn't matter.  400 datasets per setting.}\label{bigrNrho}
\begin{center}
\begin{tabular}{c|cccc|cc|rrrr}
\multicolumn{11}{l}{{\bf Experiment G}} \\
    &     &    &     &      & \multicolumn{2}{|c|}{predictor} & \multicolumn{4}{|c}{\% with $\hat{\rho}$} \\ 
setting & $N$   & $s$  & $\rho$ & $r$    & -1     &  +1    & -1 & +1 & NaN & Bad     \\ \hline
  1 & 500  & 21 & -0.95 &   53 &     1.00 & 38.65 &    45 &   0 &   7 &   52 \\
  2 & 500  & 21 & -0.50 &   53 &     9.96 & 29.78 &    10 &   0 &  10 &   20 \\
  3 & 500  & 21 &  0.00 &   53 &    19.89 & 19.89 &     3 &   1 &   1 &    5 \\
  4 & 500  & 21 &  0.50 &   53 &    29.78 &  9.96 &     0 &  12 &   1 &   13 \\
  5 & 500  & 21 &  0.95 &   53 &    38.65 &  1.00 &     0 &  47 &   1 &   48 \\
\multicolumn{11}{c}{} \\ 
  6 & 500  & 21 & -0.95 &  271 &     0.05 &  1.94 &    45 &   8 &  11 &   64 \\
  7 & 500  & 21 & -0.50 &  271 &     0.50 &  1.50 &    30 &  17 &  15 &   62 \\
  8 & 500  & 21 &  0.00 &  271 &     1.00 &  1.00 &    23 &  27 &   9 &   58 \\
  9 & 500  & 21 &  0.50 &  271 &     1.50 &  0.50 &    13 &  40 &   8 &   61 \\
 10 & 500  & 21 &  0.95 &  271 &     1.94 &  0.05 &     6 &  52 &   5 &   64 \\
\multicolumn{11}{c}{} \\ 
 11 & 500  & 21 & -0.95 & 3000 &   4.4e-4 & 1.7e-2&    23 &  30 &  31 &   83 \\
 12 & 500  & 21 & -0.50 & 3000 &   4.4e-3 & 1.3e-2&    21 &  33 &  30 &   84 \\
 13 & 500  & 21 &  0.00 & 3000 &   8.9e-3 & 8.9e-3&    22 &  31 &  31 &   83 \\
 14 & 500  & 21 &  0.50 & 3000 &   1.3e-2 & 4.4e-3&    22 &  34 &  29 &   84 \\
 15 & 500  & 21 &  0.95 & 3000 &   1.7e-2 & 4.4e-4&    20 &  35 &  25 &   80 \\
\multicolumn{11}{c}{} \\ 
 16 & 500  & 21 & -0.95 & 1e+5 &   4.0e-7 & 1.6e-5&    21 &  31 &  33 &   85 \\
 17 & 500  & 21 & -0.50 & 1e+5 &   4.0e-6 & 1.2e-5&    22 &  35 &  27 &   84 \\
 18 & 500  & 21 &  0.00 & 1e+5 &   8.0e-6 & 8.0e-6&    18 &  32 &  32 &   82 \\
 19 & 500  & 21 &  0.50 & 1e+5 &   1.2e-5 & 4.0e-6&    23 &  36 &  31 &   89 \\
 20 & 500  & 21 &  0.95 & 1e+5 &   1.6e-5 & 4.0e-7&    21 &  33 &  31 &   85 \\  \hline
\end{tabular}
\end{center}
\end{table}

Experiment G's results also show the oddity noted about Experiments A and B:  When $r$ is large, so we would expect $\hat{\rho} = -1$ and $\hat{\rho} = +1$ to occur about equally often, in fact $\hat{\rho} = +1$ is rather more frequent.  In Experiment G's settings 11 through 20, $\hat{\rho} = +1$ occurs not quite 1.5 times as often as $\hat{\rho} = -1$, even when the true $\rho$ is $-0.95$.  Section~\ref{moreDisc} discusses this further.

\subsection{The intellectual content of simulation experiments}

In the United States, most health-care research funding comes from the National Institutes of Health (NIH) and junior faculty are routinely coached about writing NIH proposals.  One maxim is that a proposal must be ``hypothesis-driven" because a proposal for descriptive research will surely fail.  We think this exclusive emphasis is mistaken;  in a new research area, descriptive research is essential (e.g., how many people, and which people, have this disease?).  We academic statisticians, however, go too far in the other direction so that our simulation experiments almost always describe operating characteristics of procedures and rarely test hypotheses.  

The distinction between hypothesis-driven and descriptive research is arguably artificial.  Does the stereotypical simulation describe or compare the new and old methods?  But posing explicit hypotheses about statistical methods --- hypotheses other than ``the new method has the correct size and higher power than the old method" --- suggests experimental designs differing from the stereotype.  As a widely-read history of molecular biology put it, ``Attractive ideas, after all, are cheap and much of the stuff of scientific genius is devising tests" (Judson 1979), for example, Meselson and Stahl's method for separating macromolecules according to buoyant density, which made it possible to verify aspects of Watson and Crick's DNA model (see also Holmes 2001).  Our own little experiments, above, are much like those of our collaborators in biology and required no great imagination;  those in Schapire (2013, 2015) required considerably more.  Perhaps we statisticians do not see the creativity in a good simulation experiment because we have made such limited use of them.  

\section{Iterate:  We haven't asked quite the right question}\label{iterate}

The hypotheses to be tested were:
\begin{enumerate}
\item As the predictor of $\hat{\rho} = -1$ becomes closer to zero, $\hat{\rho}$ is more likely to be $-1$;  as the predictor becomes larger, $\hat{\rho}$ is less likely to be $-1$. 
\item Given $N$, $s$, and $\rho$, as $r$ increases, $\hat{\rho}$ is more likely to be $-1$.
\item Given $\rho$ and $r$, as either $N$ or $s$ increases, $\hat{\rho}$ is less likely to be $-1$.
\item Given $N$, $s$, and $r$, as $\rho$ approaches $-1$, $\hat{\rho}$ is more likely to be $-1$.  
\item The effect of an increase or decrease in $r$ can be countered by an increase or decrease (respectively) in $N$ or $s$.  
\item Multiplying or dividing $r$ by about 2.5 is countered by multiplying or dividing (respectively) $N$ by a factor of about 5 or $s$ by a factor of about 3. 
\item Changes in $\rho$ have a smaller effect on the chance that $\hat{\rho} = -1$ than do changes in $N$ or $s$.  In particular, if $r$ is large enough, for any $\rho$, $\hat{\rho}$ is very likely to be $\pm 1$.  
\end{enumerate}

No result in Section~\ref{symExpts} contradicts any of Hypotheses 1 through 4, though for large enough $r$, the true value of $\rho$ doesn't matter, in the range tested.  Experiments C, D, E, and F are consistent with Hypotheses 5 and 6.  The experimental results generally are consistent with the second part of Hypothesis 7 (``if $r$ is large enough \dots").  As for the first part of Hypothesis 7 --- ``Changes in $\rho$ have a smaller effect on the chance that $\hat{\rho} = -1$ than do changes in $N$ or $s$" --- the experimental results can be interpreted as meaning ``you asked the question poorly".  With further thought, it seems that the question that motivated this inquiry --- what conditions make it likely that $\hat{\rho} = -1$? --- was a red herring.  

Experiments C through F were designed to examine how $N$, $s$, $\rho$, and $r$ affect the chance that $\hat{\rho} = -1$, and setting 5 in these experiments stands out because it is the only setting with much chance that $\hat{\rho} = +1$ or NaN (i.e., $\hat{\sigma}^2_c = 0$ or $\hat{\sigma}^2_s = 0$).  However, in experiments A, B, and G, with large values of $r$, $\hat{\rho} = -1$ was less likely than either $\hat{\rho} = +1$ or NaN.  Reducing the resolution of a design, in the sense of increasing $\sigma^2_e$, increases the chance of {\it all} kinds of bad estimate, not just $\hat{\rho} = -1$.  The disease, it seems, is poor resolution;  the different kinds of bad estimate are merely different symptoms and all improve with the same treatment, i.e., larger $s$ or $N$ or smaller $\sigma^2_e$.  

This, and the fact that Section~\ref{genHyp}'s predictor has fulfilled its purpose and can now be retired, prompted a final simulation experiment to estimate the probability of a bad estimate as a function of $N$, $s$, $\rho$, and $r$.  As in Section~\ref{genHyp}, we considered all combinations of $N \in \{50, 150, 250, \dots, 1050\}$, $s \in \{5, 15, 25, \dots, 105\}$, $\rho \in \{-0.9, -0.8, \dots, -0.1\}$, but for this final experiment $\log_{10}r$ took values in $\{0, 0.4, 0.8, \dots, 4\}$.  (It turns out that the values of $\log_{10}r$ considered in Section~\ref{genHyp} were small enough that few combinations of $N$, $s$, $\rho$, and $r$ had substantial probability of a bad estimate.  This may explain some of Section~\ref{symExpts}'s results regarding $\rho$.)  For each of the resulting $11 \times 11 \times 9 \times 11 = 11,979$ settings, we simulated and analyzed 40 datasets using the lmer function, as in Section~\ref{symExpts}.  

\begin{figure}
\caption{Percent datasets giving bad estimates as a function of $\log_{10}r$.}\label{badests}
\includegraphics[width=6.3in]{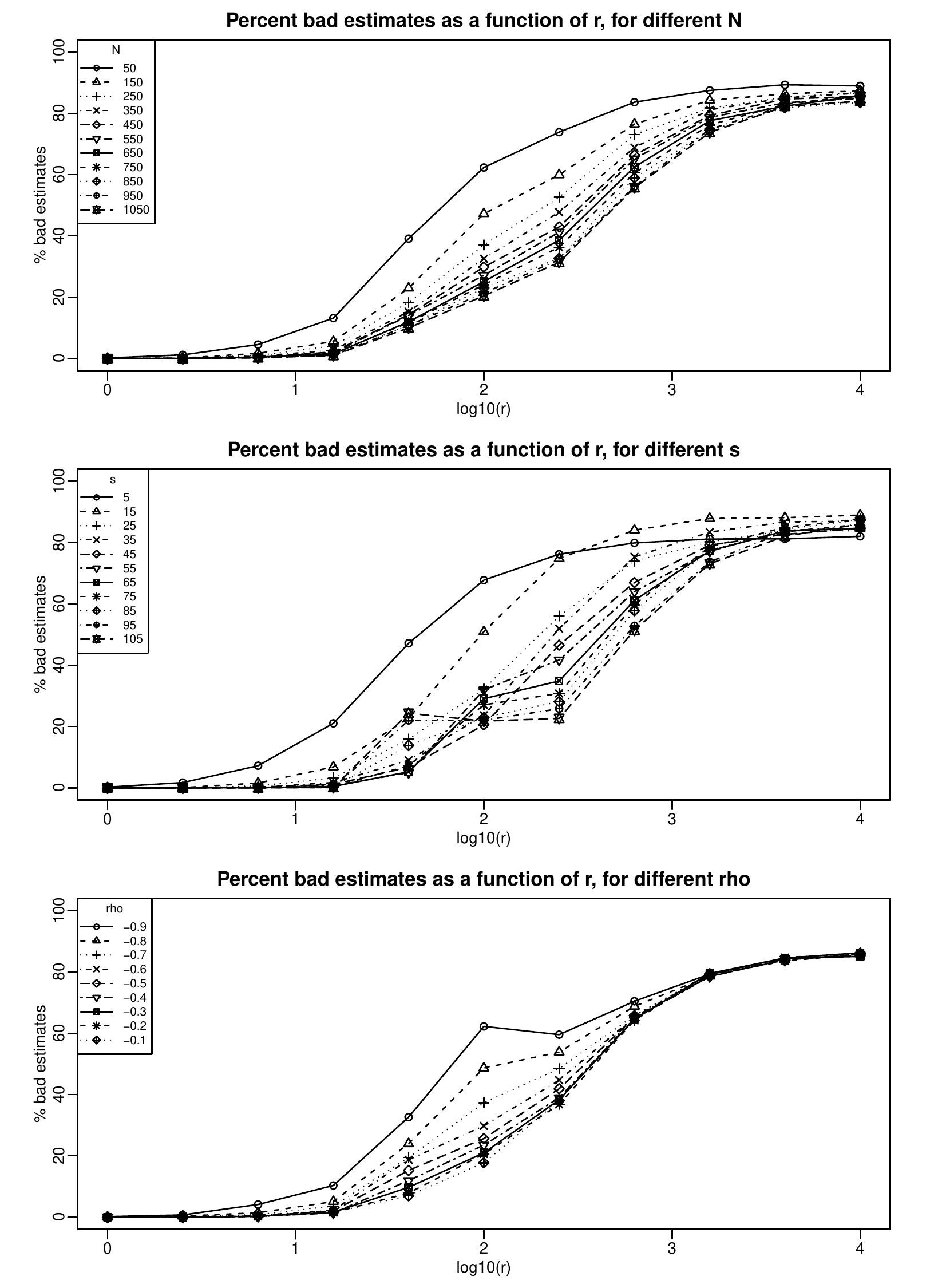}
\end{figure}

For the binary outcome ``bad estimate?~(yes/no)", treating each of $N$, $s$, $\rho$, and $r$ as a categorical factor, the four main effects and three two-way interactions involving $r$ are far larger than all the other interactions.  Figure~\ref{badests} shows interaction plots for the two-way interactions involving $r$.  In each panel, the horizontal axis is $\log_{10}r$, the vertical axis is the percent of simulated datasets giving a bad estimate, and from top to bottom, the panels show a separate line for each level of $N$, $s$, and $\rho$ respectively.  Each plotted point is the percent of $11 \times 9 \times 40 = 3960$ simulated datasets in the plots for $N$ and $s$, and of $11 \times 11 \times 40 = 4840$ simulated datasets in the plot for $\rho$.  Each plotted point has Monte Carlo standard error less than 0.8 percentage points.

When $r$ is large or small enough, the other factors have little effect on the chance of a bad estimate.  For middling $r$, $N$'s effect is simple (Figure~\ref{badests} top):  increasing $N$ reduces the chance of a bad estimate.  The same is true of $\rho$ (Figure~\ref{badests} bottom), though oddly $r$'s effect is not monotonic for $\rho = -0.9$.  Figure~\ref{badests}'s middle panel has a surprise:  For $\log_{10} r$ = 1.6 and 2.0, the effect of $s$ is not monotonic, nor is it for large $r$.  If we consider $\hat{\rho} = \pm1$ and $\hat{\rho} =$ NaN (i.e., $\hat{\sigma}^2_c = 0$ or $\hat{\sigma}^2_s = 0$) as distinct outcomes, plots analogous to Figure~\ref{badests} (in the Supplement) show that the irregularities in Figure~\ref{badests} middle arise almost entirely from $\hat{\rho} = $ NaN, while $\hat{\rho} = \pm1$ behave much more regularly.  This is a kind of competing risks situation in that a given dataset can have only one of four possible outcomes:  a ``good" estimate, $\hat{\rho} = -1$, $\hat{\rho} = +1$, or $\hat{\rho} = $ NaN;  $\hat{\rho} = $ NaN behaves most oddly as a function of $N$, $s$, $\rho$, and $r$ and because these outcomes compete, this induces oddities in results for the other outcomes.

As in the biological research we are mimicking, rich results lead to new questions.  For the present purpose of demonstration, this is a good place to stop.  

In the theory used to teach us, a hypothesis is stated and tested and the story ends.  This may be a useful way to formulate math problems for developing statistical tools but it does not describe scientific practice.  Experiments designed to test particular hypotheses often motivate a reformulation of those hypotheses or a larger structure of hypotheses and this reformulation may be the most important result of a collection of experiments.  Perhaps it is not appropriate for a scientific report to show the intermediate steps leading to that all-important reformulation, but the present paper is intended to demonstrate an approach to statistical methods research, not to present a model for scientific reports arising from it.  

\section{An ``{\it in vivo}" experiment}\label{inVivo}

The last step in many molecular-biology projects is to test all or part of the studied effect in a larger model, moving from cell cultures to mice or from rodents to larger mammals.  The object is to see if the effects found in the simple model system can be reproduced in a more complex organism.  In the random-regressions example, the analog is to see if effects from the experiments can be reproduced in a less constrained situation.  The following ``{\it in vivo}" experiment fits a model to a real dataset and then constructs artificial datasets from the fit by inflating the estimated errors $\hat{\epsilon}_{ij}$, to see whether inflating the error variance (increasing $r$) produces bad estimates as it did in the simple model system.  

The dataset is the HMO data analyzed in Hodges (1998) and re-analyzed by Wakefield (1998) and Davison (1998), available at http://www.biostat.umn.edu/$\sim$hodges/RPLMBook/\linebreak
Datasets/09\_HMO\_premiums/Ex9.html.  The outcome $y_{ij}$ is the individual health-plan premium for plan $j$ in state $i$, for $i = 1, \dots, 45$ and $j = 1, \dots, n_i$, where the $n_i$ are described below.  The total number of plans is $\sum_i n_i = 341$.  Filling the role of $x_{ij}$ is the common logarithm of the number of families enrolled in plan $(i,j)$.  The model fit to the data is:
\begin{equation}\label{inVmodel}
\begin{array}{ll}
y_{ij} =& b_0 + b_1 (\log_{10}\mbox{families enrolled)}_{ij}  + \mbox{ } [1, (\log_{10}\mbox{families enrolled)}_{ij}] (u_{i0}, u_{i1})'   \\
        &+ \mbox{ } b_{0E} \mbox{ (average expenses per admission)}_{i} + b_{0N} \mbox{(New England indicator)}_{i} + \epsilon_{ij} \nonumber
\end{array}         
\end{equation}

\noindent where $(b_0, b_1)$, $(u_{i0}, u_{i1})$, and $\epsilon_{i,j}$ are defined as above, $\mbox{(average expenses per admission)}_{i}$ is state $i$'s average expenses per hospital admission, $\mbox{(New England indicator)}_{i}$ indicates whether state $i$ is in the New England region, and $b_{0E}$ and $b_{0N}$ are scalar coefficients.  The first row of (\ref{inVmodel}) contains the population-average intercept and slope in log families enrolled and the corresponding bivariate random effect.  Except for the intercept, right-hand-side variables are standardized:  the plan-level variable ``$\log_{10}$ families enrolled" was centered and scaled using the average and standard deviation across the 341 plans;  the state-level variables were centered and scaled using the average and standard deviation across states.  (Before standardizing, the New England indicator was coded as +1 for states in New England and $-1$ for other states.)  A fit using the function lmer gave point estimates $(\hat{b}_0, \hat{b}_1, \hat{b}_{0E},\hat{b}_{0N}) = (180, -2.21, 4.78, 16.1)$ and $(\hat{\rho}, \hat{\sigma}^2_e, \hat{\sigma}^2_c, \hat{\sigma}^2_s) = (0.115, 487, 97.7, 5.39)$.  

This model and dataset differ in several ways from the model system in earlier sections: 
\begin{itemize} 
\item The fit has non-zero estimates for $(b_0, b_1)$ and has other fixed effects.
\item Within-state sample sizes vary:  $n_i$ ranges from 1 to 31 with median 5 and average 7.6.
\item The state-specific design matrices differ between states with no particular pattern.
\item The regressor ``$\log_{10}$ families enrolled" was not scaled to make $\sigma^2_c \approx \sigma^2_s$;  $\hat{\sigma}^2_c$ and $\hat{\sigma}^2_s$ differ by more than an order of magnitude.
\item The analyses cited above suggest that $\epsilon_{ij}$ is modestly right-skewed and has a higher variance in a few states.  
\end{itemize}

Artificial datasets were constructed by taking the fit to the actual data, multiplying the residuals from that fit by $\phi \ge 1$, and adding the inflated residuals to the fit.  The artificial datum $y(\phi)_{ij}$ was defined as
\begin{equation}
\begin{array}{rl}
y(\phi)_{ij} = & \mbox{fit}_{ij} + \phi \hat{\epsilon}_{ij}  \\
\mbox{where} \quad \hat{\epsilon}_{ij} = & y_{ij} - \mbox{fit}_{ij} \\
\mbox{and} \quad \mbox{fit}_{ij} = &  \hat{b}_0 + \hat{b}_1 (\log_{10} \mbox{families enrolled)}_{ij}   + \mbox{ }  [1, (\log_{10} \mbox{families enrolled})_{ij}] (\hat{u}_{i0}, \hat{u}_{i1})'   \\
        &+ \mbox{ }  \hat{b}_{0E} \mbox{ (average expenses per admission)}_{i} + \hat{b}_{0N} \mbox{ (New England indicator)}_{i},
\end{array} 
\end{equation}

\noindent where $(\hat{u}_{i0}, \hat{u}_{i1})$ are the EBLUPs computed by lmer.  Thus $\phi = 1$ gives the actual data, while $\phi > 1$ gives artificial data with inflated errors.  

Table~\ref{inVresults} shows the results of analyses for $\phi$ between 1.0 and 2.5.  As the error variance increases, the estimates become bad:  $\hat{\rho}$ increases from 0.12 in the real data to 1.00 when $\phi = 1.7$;  when $\phi$ reaches 2.4, $\hat{\sigma}^2_c$ goes to zero and $\hat{\rho}$ becomes NaN (not a number).  As Section~\ref{iterate} argued, the problem is poor resolution in the study's design;  $\hat{\rho} = -1$ and $\hat{\sigma}^2_c = 0$ are just different symptoms.  

\begin{table}
\caption{``{\it In vivo}" simulation experiment:  Increasing the error variance produces a bad estimate;  increasing it more produces a different kind of bad estimate.}\label{inVresults}
\begin{center}
\begin{tabular}{|c|r|rr|rr|}
\hline
$\phi$ &  $\hat{\rho}$  & $\hat{\sigma}^2_e$ & $\hat{\sigma}^2_e/\phi^2$&   $\hat{\sigma}^2_c$  &    $\hat{\sigma}^2_s$  \\ \hline
 1.0 & 0.115 &  487  &   487  &  97.73 & 5.39 \\
 1.1 & 0.164 &  590  &   488  &  94.13 & 5.25 \\
 1.2 & 0.230 &  704  &   489  &  88.99 & 4.97 \\
 1.3 & 0.320 &  828  &   490  &  82.36 & 4.55 \\
 1.4 & 0.444 &  963  &   491  &  74.32 & 3.98 \\
 1.5 & 0.626 & 1108  &   492  &  65.05 & 3.26 \\
 1.6 & 0.920 & 1263  &   493  &  54.77 & 2.39 \\
 1.7 & 1.000 & 1427  &   494  &  44.68 & 2.68 \\
 1.8 & 1.000 & 1599  &   494  &  34.73 & 3.21 \\
 1.9 & 1.000 & 1781  &   493  &  25.14 & 3.58 \\
 2.0 & 1.000 & 1972  &   493  &  16.36 & 3.62 \\
 2.1 & 1.000 & 2171  &   492  &   8.97 & 3.16 \\
 2.2 & 1.000 & 2379  &   492  &   3.54 & 2.02 \\
 2.3 & 1.000 & 2594  &   490  &   0.23 & 0.22 \\
 2.4 &   NaN & 2814  &   489  &   0    & 5$\times 10^{-12}$  \\
 2.5 &   NaN & 3043  &   487  &   0    & 4$\times 10^{-13}$  \\ \hline
\end{tabular}
\end{center}
\end{table}

This last step's intellectual content lies in its manner of loosening of the model system's constraints and the clarity with which it does or does not reproduce the effects found in the model system.  This step is convincing if the ``{\it in vivo}" experiment is closer to real applications and the effects of interest are demonstrated transparently.  In other words, here too much of the stuff of genius is in the design.

\section{Conclusions and discussion}\label{moreDisc}

We began by focusing on the inconvenient estimates $\hat{\rho} = \pm 1$ and learned that they are a symptom of inadequate resolution, large error variation not suppressed by sample size, with other symptoms being $\hat{\sigma}^2_c = 0$ and $\hat{\sigma}^2_s = 0$.  In that respect, the present results extend results about zero variance estimates in mixed linear models, noted in Section~\ref{genHyp}.  Although the effect of $r$, that is, $\sigma^2_e$, dominates in the sense that if $\sigma^2_e$ is small or large enough, the number of clusters and within-cluster sample size don't matter (within plausible limits), these two sample sizes {\it do} matter when $\sigma^2_e$ has a middling value.  Broadly, increasing $N$ or $s$ reduces the chance of an inconvenient estimate, and increasing $s$ has a greater effect than increasing $N$.  The implication for experimental design is that to avoid bad estimates, all else being equal it is more efficient to increase the within-cluster sample size than the number of clusters.  These results also imply that a bad estimate suggests the random-effect variance is small relative to the error variance, so it may be worthwhile to consider a model without the random effect.

Section~\ref{symExpts} left us with a puzzle, the excess of $\hat{\rho} = +1$ over $\hat{\rho} = -1$ when resolution is so poor that one might expect the two to occur equally often.  We see two possible explanations:  it is an artifact arising from the model specification or from the software.  Based on examining the log RL for many artificial datasets, when $r$ is large, the log restricted likelihood is quite flat over a large region near the maximum.  It may be that too often the model specification artifactually places a maximum at $\hat{\rho} = +1$ or that the software artifactually finds a maximum there but in either case, the restricted likelihood at $\hat{\rho} = +1$ is microscopically higher than at all other points in a large region and it would be helpful if software reported that.  

As for how we statisticians learn about our methods, the example shows a few things.  First, results like this could never be discovered using asymptotic methods because the fundamental problem is insufficient information and at the asymptote, we have infinite information.  It's also hard to imagine how the previous paragraph's puzzle could be detected except in simulation experiments.  Second, we produced useful facts with math and computing exercises that could be executed by a capable Master's student under faculty supervision.  Nonetheless, the results are useful and the design of each step in the process can have substantial intellectual content, though we make no grand claims about the present paper's designs.  If utility has merit, then these contributions imply that excellent empirical studies of statistical methods merit publication as much as theorems.

\vspace{20pt}
{\centerline{\large \bf Acknowledgements}}

In writing this paper I had the benefit of comments and suggestions by Ning Dai, Michael Lavine, and Wei Pan. Birgit Grund suggested that a bad estimate might be considered a signal to omit the random effect from the model;  Weihua Guan suggested the alternate version of Experiments A and B in which $\sigma^2_e$ was fixed.  I especially thank Patrick Schnell for reading drafts carefully and making many great suggestions, including Section 1's argument that $\hat{\rho} = \pm 1$ is a problem and Figure 1.  These generous people do not necessarily agree with the views expressed in this paper.

\vspace{20pt}
{\centerline{\large \bf References}}
\begin{description}

\item Adams JL (1990a).  Evaluating regression strategies.  PhD dissertation, University of Minnesota School of Statistics.  

\item Adams JL (1990b).  A Computer Experiment to Evaluate Regression Strategies. {\it Proceedings of the American Statistical Association}, 1990:55--62.    

\item Bates D, Maechler M, Bolker B, Walker S, Christensen RHB, Singmann H, Dai B (2014).  R package lme4.  URL https://cran.r-project.org/web/packages/lme4/index.html.

\item Clifton K (1997).  An empirical assessment of the normal approximations for logistic regression.  Unpublished MS thesis, Division of Biostatistics, University of Minnesota.  

\item Davison AC (1998).  Discussion of Hodges (1998).  {\it J.~Royal Stat.~Soc., Series B}, {\bf60}:529-530.

\item Feyerabend P (1993).  \textit{Against Method}.  New York:Verso.

\item Friedman J, Hastie T, Tibshirani R (2000).  Additive logistic regression:  A statistical view of boosting (with discussion).  {\it Ann.~Stat.}, {\bf 28}:337-407.  

\item Hill BM (1965).  Inference about variance components in the one-way model.  \textit{J.~American Stat.~Assn.}, {\bf 60}:806-825

\item Hodges JS (1998).  Some algebra and geometry for hierarchical models, applied to diagnostics (with discussion).  {\it J.~Royal Stat.~Soc., Series B}, {\bf 60}:497--536.

\item Hodges JS (2014).  \textit{Richly Parameterized Linear Models}.  Boca Raton, FL:  Chapman \& Hall.  

\item Holmes FL (2001).  {\it Meselson, Stahl, and the Replication of DNA. A History of ``The Most Beautiful Experiment in Biology"}.  New Haven:  Yale University Press.

\item Huppler Hullsiek, K (1996).  Assessing the accuracy of normal approximations from proportional hazards regression.  Unpublished MS thesis, Division of Biostatistics, University of Minnesota.  

\item Judson HF (1979).  \textit{The Eighth Day of Creation}.  New York:  Simon \& Schuster.

\item Larntz K (1978).  Small-sample comparisons of exact levels for chi-squared goodness-of-fit statistics. \textit{J.~American Stat.~Assn.}, {\bf 73}, pp. 253--263.

\item R Core Team (2014). R: A language and environment for statistical computing. R Foundation for Statistical Computing, Vienna, Austria. URL http://www.R-project.org/.

\item Ruppert D, Wand MP, Carroll RJ (2003).  \textit{Semiparametric Regression}.  New York:  Cambridge University Press. 

\item Schapire RE (2013).  Explaining AdaBoost. In Bernhard Sch\"{o}lkopf, Zhiyuan Luo, Vladimir Vovk, editors, {\it Empirical Inference: Festschrift in Honor of Vladimir N. Vapnik}, New York:Springer, 37--52.

\item Schapire RE (2015).  Explaining AdaBoost.  Joint Statistical Meetings 2015, Seattle Washington, abstract at URL http://www.amstat.org/meetings/JSM/2015/onlineprogram/\newline AbstractDetails.cfm?abstractid=317916
 
\item Wakefield J (1998).  Discussion of Hodges (1998).  {\it J.~Royal Stat.~Soc., Series B}, {\bf60}:523--526, with figures on pp.~526-529.

\end{description}

\end{document}